\documentclass[10pt,conference]{IEEEtran}
\IEEEoverridecommandlockouts

\def\BibTeX{{\rm B\kern-.05em{\sc i\kern-.025em b}\kern-.08em
    T\kern-.1667em\lower.7ex\hbox{E}\kern-.125emX}}

\usepackage{amsmath,amssymb,amsfonts}
\usepackage{algorithmic}
\usepackage{graphicx}
\usepackage{textcomp}
\usepackage{cite}
\usepackage{tabularx}
\usepackage{enumitem}
\usepackage{array}
\usepackage{booktabs}
\usepackage[table]{xcolor}  
\usepackage{tcolorbox}      
\usepackage{url}
\usepackage{setspace}

\usepackage{hyperref}       
\begin{document}

\title{An Empirical Validation  of Open Source Repository Stability Metrics\\}

\author{\IEEEauthorblockN{1\textsuperscript{st} Elijah Kayode Adejumo}
\IEEEauthorblockA{\textit{Computer Science} \\
\textit{George Mason University}\\
Fairfax, USA \\
eadejumo@gmu.edu}
\and
\IEEEauthorblockN{2\textsuperscript{nd} Brittany Johnson}
\IEEEauthorblockA{\textit{Computer Science} \\
\textit{George Mason University}\\
Fairfax, USA \\
johnsonb@gmu.edu}}

\maketitle

\begin{abstract}
Over the past few decades, open source software has been continuously integrated into software supply chains worldwide, drastically increasing reliance and dependence. Because of the role this software plays, it is important to understand ways to measure and promote its stability and potential for sustainability. 
Recent work proposed the use of control theory to understand repository stability and evaluate repositories' ability to return to equilibrium after a disturbance such as the introduction of a new feature request, a spike in bug reports, or even the influx or departure of contributors.
This approach leverages commit frequency patterns, issue resolution rate, pull request merge rate, and community activity engagement to provide a Composite Stability Index (CSI). 
While this framework has theoretical foundations, there is no empirical validation of the CSI  in practice. In this paper, we present the first empirical validation of the proposed CSI by experimenting with 100 highly ranked GitHub repositories. 
Our results suggest that (1) sampling weekly commit frequency pattern instead of daily is a more feasible measure of commit frequency stability across repositories and (2) improved statistical inferences (swapping mean with median), particularly with ascertaining resolution and review times in issues and pull request, improves the overall issue and pull request stability index. Drawing on our empirical dataset, we also derive data‐driven half‐width parameters that better align stability scores with real project behavior. These findings both confirm the viability of a control-theoretic lens on open-source health and provide concrete, evidence-backed applications for real-world project monitoring tools.

\end{abstract}


\section{Introduction}
Since the early introduction and broad acceptance of open source software \cite{fuggetta2003open}, concerns have been raised regarding the long term sustainability  \cite{chang2007open, chengalur2010sustainability, nyman2013code} and stability \cite{bouktif2014predicting,raemaekers2012measuring} of open source innovation. 
Numerous techniques and suggestions have been proposed to effectively maintain open source projects~\cite{aberdour2007achieving, fogel2005producing}. 
There have also been numerous efforts to understand practical and actionable metrics\cite{germonprez2017theory,germonprez2018eight} that support improved engagement to open source software. These metrics are especially beneficial to community managers (maintainers).  
With the rapid integration of open source software into diverse workflows \cite{hauge2010adoption}, it is imperative to evaluate all factors and metrics that contribute to its sustainability and stability. 

Most prior work aimed at supporting open source sustainability has focused on facilitating  continuous contributions. Many research efforts with this goal have focused on investigating and evaluating tool support for  open source on-boarding process~\cite{steinmacher2015supporting,balali2020recommending}.
Tool support for on-boarding newcomers ranges from various kinds of recommendation systems and mentorship matching to gamifying contributions and interactions~\cite{santos2024software}.
A subset of the prior work on tool support has focused on detecting and recommending good first issues for newcomers~\cite{stanik2018simple, horiguchi2021onboarding,Steinmacher4}.
Recently, in light of the current trends, large language models (LLMs) have been proposed as a means to enhance the on-boarding experience through various documentation transformation techniques ~\cite{adejumo2024towards, tufano2024unveiling}. 
More broadly, there have been numerous studies on understanding trends in open source communities \cite{martinez2014current}, including research on learning culture, collective intelligence, motivation, innovation, community structure, success, and virtual organization. These studies have explored necessary support mechanisms through knowledge sharing techniques and the utilization of social Q\&A websites~\cite{sharma2002framework,vasilescu2014social} for open source software communities.

As suggested by prior work, livelihood and survival of the open source ecosystem is a critical concern for all participants, including software developers, end-users, and investors. Participants require sufficient knowledge about ecosystem health and wellness, which in turn builds trust and attracts greater recognition and investment \cite{jansen2014measuring}. 
Static repository health metrics provided by tools like CHAOSS \cite{goggins2021open} offer maintainers access to various metrics such as the ratio of new and returning contributors, new downloads, knowledge and artifact creation patterns, mailing list responsiveness, bug fix time, new patents, usage patterns, and variations in contributor types. These metrics provide valuable insights into repository health.

However, open source software systems are inherently dynamic \cite{wu2006open,vasa2010growth} and are often influenced by social behavior \cite{yin2022open}, making it necessary to consider assessment or evaluation from a dynamic perspective. 
Recently, Destefanis et al. \cite{destefanis2025introducing} introduced the Composite Stability Index (CSI) framework, which utilizes control theory dynamics to ascertain repository stability—that is, repositories' ability to return to equilibrium after disruptions. The framework considers key factors such as commit frequency, issue resolution, pull request merging, and community engagement patterns. However, its practicality and feasibility have not yet been determined.

To this end, we carefully curated a dataset of 100 repositories to investigate use of the CSI in practice. We found that contrary to the daily contribution frequency pattern suggested in the CSI framework, weekly aggregation proves more practical among projects. We also found that the  statistical inferences (mean) applied by the original CSI could be improved for more practical application and feasibility, particularly regarding issue resolution and pull request review times. Finally, based on our empirical evaluations, we derived and recommend a data-driven half-width parameter for the triangular normalizer, which has a significant influence on CSI variance.


In the following sections~\ref{sec:Background_RelatedWork}--\ref{sec:Threat_to_validity}, we present the related work, methodology, results, and discussion.

\section{Background \& Related Work}
\label{sec:Background_RelatedWork}

To ground our empirical investigation, we first survey prior efforts on repository-level metrics, repository health, and theoretical stability models. We then focus on recent control-theoretic approaches to assessing repository stability, particularly the Composite Stability Index (CSI), and address the empirical validation on real-world projects. 

\subsection{Repository Level Metrics}
Advancements in software engineering, both in industry and research, have been driven by the ability to obtain and curate data from software repositories~\cite{6224294}. 
GitHub's REST API enables researchers to extract meaningful data and uncover hidden patterns in software development, leading to significant advances in software engineering \cite{kalliamvakou2014promises}. \textbf{Commit history}, the evolutionary record of changes to source code and other artifacts in version control systems like GitHub \cite{blischak2016quick} has been instrumental in revealing various patterns. These patterns have proven valuable for predicting potential fault-proneness of classes in object-oriented software \cite{chong2018can}, analyzing developer commit patterns as indicators of code quality and developer expertize \cite{6895411}, and conducting commit impact analysis to understand software quality evolution \cite{8009930}. Numerous other studies have demonstrated how important commit history metrics has been advancing software engineering research~\cite{weicheng2013mining,tsantalis2018accurate,eliseeva2023commit}.

GitHub utilizes \textbf{pull requests} to facilitate collaborative development among developers working on or interested in a project. This approach allows developers to work on project features independently without committing every individual change directly to the main codebase, thereby mitigating frequent merge conflicts among team members. 
Once a substantial set of changes has been completed, the developer creates a pull request, which is then reviewed by maintainers or project owners/administrators who can approve and merge the commits \cite{10.1145/2597073.2597121}.
The significance of pull-based software development \cite{10.1145/2568225.2568260} has attracted considerable research attention, particularly in developing efficient methods for assigning appropriate reviewers to review pull requests \cite{6976151}. Additionally, several studies have proposed strategies for how developers and teams can maximize the benefits of pull requests \cite{jiang2021recommending,batoun2023empirical,wessel2023github}.  

Faults discovered by developers or users of a software system are reported to maintainers or administrators through \textbf{issue tracking} systems such as Jira and Bugzilla. Open source software, particularly projects hosted on GitHub, utilize built-in issue reporting and tracking systems. These systems have significantly improved the reporting, management, and resolution of software issues \cite{6698918}. Extensive research has been conducted to enhance user experiences in issue tracking systems and to develop tools and techniques that support better reporting, management, and resolution of issue reports~\cite{bettenburg2008makes,sun2011towards,tian2012improved}.

\textbf{Comments} on commits, issues, and pull requests play a vital role in revealing communication patterns and the overall health of open source software projects. 
Prior studies have proposed using comments to understand the emotions of users and contributors in open source software development~\cite{10.1145/2597073.2597118, 10.1145/3194932.3194936}. GitHub discussions have also proven effective in helping project communities address errors and unexpected behaviors, ultimately advancing project development \cite{hata2022github}.
The trends and patterns derived from these repository metrics have proven valuable in addressing problems and proposing diverse solutions within open source software development and software engineering research more broadly.

\subsection{Repository Health and Sustainability}
The health and sustainability of open source projects shows provides insight to the reliability of the project which is a  critical factors that consumers and contributors  consider before adopting a project as a dependency or integrating into workflow \cite{sanchez2020open} 
The CHAOSS initiative~\cite{goggins2021open} provides a curated, visualizable dataset of health metrics such as conversion rate, issue age, change request closure ratio and many other metrics that are useful for determining project health.
Crowston et al. recommended that consumers also examine community engagement patterns through project websites, mailing lists, and list archives before choosing to adopt a particular open source project~\cite{crowston2006assessing}. A study by Schweik~\cite{schweik2013sustainability} found that developers are motivated by multiple factors such as learning opportunities, incentives, contributing to the public good, or genuine passion for the project, all of which could be crucial to project sustainability.
Several other studies have proposed techniques and revealed significant findings regarding how sustainability can be achieved in open source software through efficient code forking and improved patch contributions \cite{sethanandha2010managing,nyman2013code,gamalielsson2014sustainability}.

\subsection{Stability in Software Engineering}
Software stability is defined as a software's resistance to ripple effects caused by modifications, bugs, and other disturbances to the software \cite{yau1980some}, 
stability in software spread across diverse areas in software engineering. 
From the perspective of software maintenance processes~\cite{yau1980some,mattsson2006software}, stability is used to understand the consequences of modifications, such as maintenance costs and potential errors that could emerge when developing maintenance plans.
Stability has also been considered in software evolution planning, where studies have identified it as an important criterion for evaluating design and making design decisions \cite{gomaa2011software,jazayeri2002architectural}. There has been considerable research into understanding how stability relates to other areas, including software aging \cite{cotroneo2014survey}, software reuse \cite{frakes2005software}, incremental software development \cite{abbas2014value}, and adaptation \cite{salehie2009self,krupitzer2015survey}.

The concept of stability in control theory centers on how systems evolve over time and react to disruptions. Stability is achieved when a system, after being displaced from its equilibrium position, demonstrates a natural tendency to restore itself to that equilibrium. Several studies have explored the application of control theory to software engineering \cite{filieri2015software,shevtsov2017control}.   
However, there was no unified framework for assessing stability, especially considering the socio-technical \cite{ducheneaut2005socialization} nature of open source software repositories.
Recently, Destefanis et al. introduced a unified framework known as the Composite Stability Index \cite{destefanis2025introducing}, which integrates key repository metrics into a weighted framework. They proposed that the Composite Stability Index consists of four major repository properties that represent the state of a repository as:
\begin{equation}
\mathbf{R}(t)=\bigl[c(t),\,i(t),\,p(t),\,a(t)\bigr]^{\mathsf T},
\label{eq:R_vector}
\end{equation}
where the properties are,  
\begin{itemize}
  \item $c(t)$: commit frequency function;
  \item $i(t)$: issue resolution rate function;
  \item $p(t)$: pull request merge rate function;
  \item $a(t)$: activity engagement function.
\end{itemize}

\par\vspace{0.7\baselineskip}
\subsubsection* {\textbf{Composite Stability Metric Definitions}} 
The commit frequency, issue resolution rate, pull request merge rate, and activity engagement represent the fundamental dimensions of repository activity and health \cite{destefanis2025introducing}.

The commit frequency function $c(t)$ provides insights into development patterns, showing the frequency of code changes and ultimately revealing how active the project is through commit frequencies and their timestamps. 
The issue resolution function $i(t)$ provides insights into how the project handles and resolves problems or concerns raised. This is calculated by analyzing issue creation and closure timestamps. 
The pull request merge rate function $p(t)$ provides insights into how the project handles pull requests; it assesses code review and integration processes by utilizing pull request creation and entire lifecycle timestamps. 
The activity engagement function $a(t)$ provides overall insights into repository engagement through comment activity and interactions. 

The authors conceptualized the repository as a dynamical system and provided detailed definitions for each component as follows:
\par\vspace{0.7\baselineskip}
\noindent\emph{Commit Frequency Function}

\begin{equation}
c(t)=\frac{N_{C}\!\bigl(t,\,t+\Delta t\bigr)}{\Delta t},
\label{eq:commit_freq}
\end{equation}

Where $N_{C}\!\bigl (t,\,t+\Delta t\bigr)$ represents the number of commit within the interval $ \!\bigl(t,\,t+\Delta t\bigr)$, this data can be curated from available commit history.

\par\vspace{0.7\baselineskip}

\noindent\emph{Issue Resolution Rate}

\begin{equation}
i(t)=
\frac{N_{i}^{\text{closed}}\!\bigl(t,\,t+\Delta t\bigr)}
     {N_{i}^{\text{total}}(t)}
\;\cdot\;
\frac{1}{1+\overline{T}_{\text{resolution}}(t)},
\label{eq:issue_resolution_rate}
\end{equation}

Where, ${N_{i}^{\text{closed}}\!\bigl(t,\,t+\Delta t\bigr)} $ represents the number of closed issues within the interval  $ \!\bigl(t,\,t+\Delta t\bigr)$, ${N_{i}^{\text{total}}(t)}$ represents the total issues and  ${\overline{T}_{\text{resolution}}(t)}$ represents the average resolution time for issues to be closed in the interval $ \!\bigl(t,\,t+\Delta t\bigr)$.
\par\vspace{0.7\baselineskip}
\noindent\emph{Pull Request Merge Rate}
\begin{equation}
p(t)=\frac{N_{p}^{\text{merged}}\!\bigl(t,\,t+\Delta t\bigr)}
     {N_{p}^{\text{total}}(t)}
\;\cdot\;
\frac{1}{1+\overline{T}_{\text{review}}(t)},
\label{eq:pr_merge_rate}
\end{equation}

Where, ${N_{p}^{\text{merged}}\!\bigl(t,\,t+\Delta t\bigr)}$ represents the number of pull request merged within the interval $ \!\bigl(t,\,t+\Delta t\bigr)$,  ${N_{p}^{\text{total}}(t)}$ represents the total pull request, while $\overline{T}_{\text{review}}(t)$ represents the average review time for pull request in the interval $ \!\bigl(t,\,t+\Delta t\bigr)$.

\par\vspace{0.7\baselineskip}
\noindent\emph{Activity Engagement Function} 
\begin{equation} 
a(t) = \frac{N_{\text{comments}}(t, t+\Delta t)}{N_{\text{issues}}(t) + N_{\text{prs}}(t)} \cdot \frac{N_{\text{active\_users}}(t, t+\Delta t)}{N_{\text{total\_users}}(t)}, 
\label{eq:activity_engagement} 
\end{equation} 
Where, $N_{\text{comments}}(t, t+\Delta t)$ represents the number of comments in the interval $(t, t+\Delta t)$, $N_{\text{issues}}(t) = N_{i}^{\text{total}}(t) - N_{i}^{\text{closed}}(t)$, and $N_{\text{prs}}(t) = N_{p}^{\text{total}}(t) - N_{p}^{\text{merged}}(t)$ represent the number of open issues and open pull requests at time $t$, respectively. $N_{\text{active\_users}}(t, t+\Delta t)$ represents users who have interacted with the repository through comments, commits, or pull requests within the interval $(t, t+\Delta t)$, and $N_{\text{total\_users}}(t)$ represents the total number of users who have interacted with the repository.

These stability metrics capture the core facets of a repository, but they require rigorous empirical evaluation to confirm their effectiveness. 
In this paper, we evaluate and discuss the effectiveness of these metrics as stability indicators through practical validation methods. 

\section{Methodology}
\label{sec:Methodology}
Our study aims to empirically evaluate how well the Composite Stability Index (CSI) framework proposed in prior work~\cite{destefanis2025introducing} applies to real-world software repositories. 
To assess the practical adequacy of the CSI equations
(Eqs.~\eqref{eq:commit_freq}–\eqref{eq:activity_engagement}), we pose the following research questions:\\

\begin{description}
    \item[\textbf{RQ$_{1}$}]To what extent do widely adopted open-source repositories satisfy the individual CSI criteria for commit frequency, issue resolution, pull-request merging, and activity engagement?
\item[\textbf{RQ$_{2}$}] How does swapping outlier-sensitive measures and high-frequency sampling for robust, window-bounded alternatives affect stability thresholds?

\item[\textbf{RQ$_{3}$}] How sensitive is the CSI classification to variations in the triangular‐normaliser parameters ($\mu_{k}$, $\sigma_{k}$) across real-world repositories?

\end{description}


\par\vspace{0.7\baselineskip}

For each research question, we specified the analysis window parameter as 5 years. That is, in interval
\(\bigl(t,\,t+\Delta t\bigr)\), \(\Delta t = 5\text{ years}\) on each of the repositories we analyzed, which we discuss next.

\subsubsection*{\textbf{Repository Selection}}
\label{sec:repo_selection}
To evaluate the practicality of the CSI, we
curated a sample of 100 GitHub repositories that satisfy five objective characteristics:

\begin{enumerate}[label=\roman*.]
  \item \textbf{Stars $>$ 10,000 }:  
        we believe a large watcher base signals sustained interest \cite{borges2018s} ensuring that activity metrics are representative of
        projects people actually rely on.
  \item \textbf{Forks $>$ 9,000 }: 
        Fork counts correlate with external contribution and merge
        traffic \cite{jiang2017and}; high values stress test the pull-request component of
        CSI.
  \item \textbf{Maturity $\ge$ 10 years old}.  
        We believe a decade of history provides the longitudinal data needed to
        observe stability trends and cushions against short-lived bursts
        of activity.
  \item \textbf{Not for education}.  
      Projects such as books, programming tutorials, and screencasts attract stars and discussion but contain relatively little code evolution. Given our focus on software development dynamics, we removed these repositories from our CSI analysis.
\item \textbf{Not archived}.  
      Any repository flagged as \emph{archived} was removed during manual
      screening so that the final dataset contains only actively maintained projects.
\end{enumerate}

We curated our dataset based on the inclusion criteria above and now detail the data extraction pipeline and the analytical procedures used to answer each research question.


\subsection{Data Extraction Pipeline}\
\textbf{Window definition}
$ (\Delta t\bigr)$ = 5 years. 

We utilized the following GitHub REST endpoints to curate relevant data:  
\textbf{Commits:} GET /repos/{o}/{r}/commits?since= 
\textbf{Issues \& PRs:} GET /search/issues 
    \textbf{Comments:} GET /issues/comments, GET/pulls/comments. 

\textbf{Caching \& retries: } we utilized 24 hour JSON cache per endpoint call. We also implemented transient \(\text{4xx}/\text{5xx}\) errors trigger exponential back-off with up to five retries. 

\textbf{Reproducibility: } All scripts, csv files and raw JSON dumps are published: https://anonymous.4open.science/r/OSS-stability-3C1F/ 

\subsection{Operationalization of the Original CSI Baseline (RQ1)}
\label{sec:operationalization of CSI Baseline}
To evaluate practically the original CSI framework, we implemented the thresholds and metrics exactly as defined in \cite{destefanis2025introducing}.  

\textbf{Commit Stability:} To determine commit stability based on the commit frequency function in Equation~\eqref{eq:commit_freq}, we evaluate stability for each of the 100 repositories in our sample dataset (Section~\ref{sec:repo_selection}). A repository's commit pattern is classified as stable when:
\begin{equation}
\left|\frac{dc(t)}{dt}\right| \;\le\; \alpha_{c},
\quad \forall\, t \in [t_{0},\, t_{0}+T],
\label{eq:commit_stability}
\end{equation}

where $\alpha_{c}$ sets an upper limit on the allowable rate of change in commit frequency. If the rate of change exceeds this threshold, the repository is no longer classified as stable. The threshold $\alpha_{c}$ is defined as:
\begin{equation}
\alpha_{c} = \frac{\sigma_{\text{daily commits}}}{\mu_{\text{daily commits}}} \le 0.5.
\label{eq:threhold limit}
\end{equation}

\textbf{Issue Management Stability:} To determine stable issue resolution patterns across our sample dataset, we first apply the issue resolution rate function \eqref{eq:issue_resolution_rate}, then utilized the issue stability threshold:
\begin{equation}
i(t) \ge \beta_{i} \;\text{and}\; \overline{T}_{\text{resolution}}(t) \le \tau_{i},
\quad \forall t \in [t_{0}, t_{0}+T],
\label{eq:issue_stability}
\end{equation}
where $\beta_{i}$ is the minimum acceptable issue resolution rate with a threshold of 0.3, and $\overline{T}_{\text{resolution}}(t)$ is the average resolution time with a maximum acceptable threshold of $\tau_{i} = 14$ days.

\textbf{Pull Request Processing Stability:}
To determine stable pull request processing patterns across our sample dataset, we first apply the pull request merge rate function \eqref{eq:pr_merge_rate}, then evaluate the pull request processing threshold:
\begin{equation}
p(t) \ge \beta_{p} \;\text{and}\; \overline{T}_{\text{review}}(t) \le \tau_{p},
\quad \forall t \in [t_{0}, t_{0}+T],
\label{eq:pr_stability}
\end{equation}
where $\beta_{p}$ is the minimum acceptable pull request merge rate with a threshold of 0.4, and $\overline{T}_{\text{review}}(t)$ is the average review time with a maximum acceptable threshold of $\tau_{p} = 5$ days.

\textbf{Community Engagement Stability:} To determine stable community activity engagement across our dataset, we first apply the activity engagement function \eqref{eq:activity_engagement}, then evaluate the community engagement stability threshold: 

\begin{equation}
a(t)\ge \gamma_{a}\;\text{and}\;
\frac{N_{\text{active\_users}}(t)}{N_{\text{total\_users}}(t)}
      \ge \delta_{a},
\quad\forall\,t\in[t_{0},\,t_{0}+T],
\label{eq:activity_stability}
\end{equation}

Where, $\gamma_{a}$ is the minimum acceptable activity ratio with a threshold of 0.25, and $\delta_{a}$ is the minimum acceptable active user ratio with a threshold of 0.15.

\subsection{Composite Stability Index (CSI)}
To aggregate the four stability dimensions into a single score, we utilize the Composite Stability Index introduced by Destefanis \textit{et al.}~\cite{destefanis2025introducing}:
\begin{equation}
\mathrm{CSI}(t) = w_{c}\,\phi_{c}(c(t)) + w_{i}\,\phi_{i}(i(t)) + w_{p}\,\phi_{p}(p(t)) + w_{a}\,\phi_{a}(a(t)),
\label{eq:csi}
\end{equation}
with the weight vector
\begin{equation}
W = [w_{c}, w_{i}, w_{p}, w_{a}] = [0.3, 0.25, 0.25, 0.2].
\label{eq:weight_vector}
\end{equation}

\textbf{Triangular Normalizer:} Each raw metric $x \in \{c, i, p, a\}$ is mapped onto the unit interval by
\begin{equation}
\phi_{k}(x) =
\begin{cases}
  1 - \frac{|x - \mu_{k}|}{\sigma_{k}}, & \text{if } |x - \mu_{k}| \le \sigma_{k}, \\[6pt]
  0, & \text{otherwise},
\end{cases}
\label{eq:triangular_normalizer}
\end{equation}
where $\mu_{k}$ is the target value and $\sigma_{k}$ is the admissible deviation for component $k$. 

Target and tolerance values from the triangular normalizer Table \ref{tab:csi-thresholds} defined by Destefanis et al. \cite{destefanis2025introducing} has implications as following: 

\begin{enumerate}[label=\roman*.]
  \item \textbf{Commit pattern (\(\phi_c\)).}  \(\mu_c = 0.25\) (i.e., we expect roughly a coefficient of variation of 0.25).  A repository’s measured \(c\) that exactly matches 0.25 yields \(\phi_c = 1\).  If \(c\) lies anywhere between 0.00 and 0.50, then 
  \[
    \phi_c(c) \;=\; 1 - \frac{\lvert\,c - 0.25\,\rvert}{0.25}.
  \]
  As soon as \(c\) exceeds 0.50 or falls below 0.00, \(\phi_c\) becomes zero.  
  In other words, if a project’s commit‐frequency CV strays beyond \(\pm0.25\) around 0.25, it no longer contributes positively to the CSI.

  \item \textbf{Issue management (\(\phi_i\)).}  Here \(\mu_i = 0.40\) (target resolution rate) and \(\sigma_i = 0.10\).  Thus:
  \[
    \phi_i(i) \;=\; 
    \begin{cases}
      1 - \tfrac{\lvert\,i - 0.40\,\rvert}{0.10}, & \text{if } 0.30 \le i \le 0.50,\\
      0, & \text{otherwise}.
    \end{cases}
  \]
  If a repository closes issues at exactly 40\% in the given interval, \(\phi_i = 1\).  If the rate dips to 30\% or climbs to 50\%, \(\phi_i\) reaches 0, because it is at the edge of the admissible band.

  \item \textbf{Pull‐request stability (\(\phi_p\)).}  With \(\mu_p = 0.50\) and \(\sigma_p = 0.10\), any merge‐rate \(p\) in \([0.40,\,0.60]\) yields a strictly positive \(\phi_p\).  For example, if \(p = 0.55\), then
  \[
    \phi_p(0.55) \;=\; 1 - \frac{\lvert\,0.55 - 0.50\,\rvert}{0.10} \;=\; 1 - 0.5 \;=\; 0.5.
  \]
  Once \(p < 0.40\) or \(p > 0.60\), \(\phi_p\) collapses to 0.

  \item \textbf{Community engagement (\(\phi_a\)).}  The  required  target activity ratio of \(\mu_a = 0.35\) and allow a deviation of \(\sigma_a = 0.10\).  Hence, \(\phi_a(a)\) is nonzero only when \(0.25 \le a \le 0.45\).  At \(a = 0.35\), \(\phi_a = 1\).  At the tolerance edges (\(a = 0.25\) or \(a = 0.45\)), \(\phi_a = 0\).
\end{enumerate}

\begin{table}[ht]
\centering
\caption{Target values ($\mu_{k}$) and tolerances ($\sigma_{k}$) used by the CSI.}
\label{tab:csi-thresholds}
\begin{tabular}{lcc}
\toprule
Component $k$ & $\mu_{k}$ & $\sigma_{k}$ \\
\midrule
Commit pattern $(\phi_{c})$          & 0.25 & 0.25 \\
Issue management $(\phi_{i})$        & 0.40 & 0.10 \\
Pull-request stability $(\phi_{p})$  & 0.50 & 0.10 \\
Community engagement $(\phi_{a})$    & 0.35 & 0.10 \\
\bottomrule
\end{tabular}
\end{table}

\subsection{Estimator and Sampling Robustness (RQ2)}
\label{sec:RQ2_methodology}
Statistical inferences play an important role in the CSI framework, as demonstrated in Equations~\eqref{eq:issue_resolution_rate} and \eqref{eq:pr_merge_rate}, where average issue resolution rates and pull request review rates are calculated. However, several studies have highlighted potential robustness concerns with these approaches,such as skewness in bug resolution~\cite{eiroa2023large} and commit intervals~\cite{kolassa2013empirical}.
In these cases, prior work recommends using medians as indicators to mitigate outliers or measurement of atypical activity.


The commit history stability index specified in Equation~\eqref{eq:commit_stability} utilizes daily commit patterns where the standard deviation remains less than half of the mean. However, daily commit patterns may be affected by timezone differences, as contributors to open source software are often distributed across continents \cite{wachs2022geography} and may have full-time software development roles \cite{alexander2002working,ghosh2005understanding}. 
Weekly aggregation of commit history patterns offers a more logical approach, as studies have shown that aggregating high-frequency data reduces noise \cite{hyndman2018forecasting}.
These considerations motivate our investigation of weekly aggregation for commit history stability thresholds and the adoption of robust statistical estimators throughout the CSI framework.

\subsection{Sensitivity Analysis of CSI Classification (RQ3)}
In each of the CSI components- commit, issues, pull request, and activity engagement-repositories must attain stability by meeting the thresholds as defined in equations \eqref{eq:commit_stability}--\eqref{eq:activity_stability} (Section~\ref{sec:operationalization of CSI Baseline}) before their stability measures are evaluated by the triangular-normalizer parameter. This ensures that the stability measures are within a target $\mu_k$ and a tolerance measure $\sigma_k$ as shown in Table~\ref{tab:csi-thresholds} before they contribute variance to the $CSI(t)$. 
However, these target values are proposed based on experience of the authors \cite{destefanis2025introducing} and have not been evaluated to determine their feasibility.
To answer \textbf{RQ$_3$}, we performed descriptive statistical analysis across repositories that meet the stability threshold in their individual CSI components but do not contribute variance to the $CSI(t)$ due to their stability measures being outside the target and tolerance measures. 
This analysis shows the practicality of the target and tolerance values proposed.

\section{Results}
\label{sec:Results}
In this section, we discuss the findings from our evaluation of the original CSI thresholds \cite{destefanis2025introducing} and the impact of threshold variance.

\subsection{Satisfaction of Original CSI Criteria (\textbf{RQ$_1$})}

\begin{figure}[ht]
		\centering
		\includegraphics[width=0.48\textwidth]{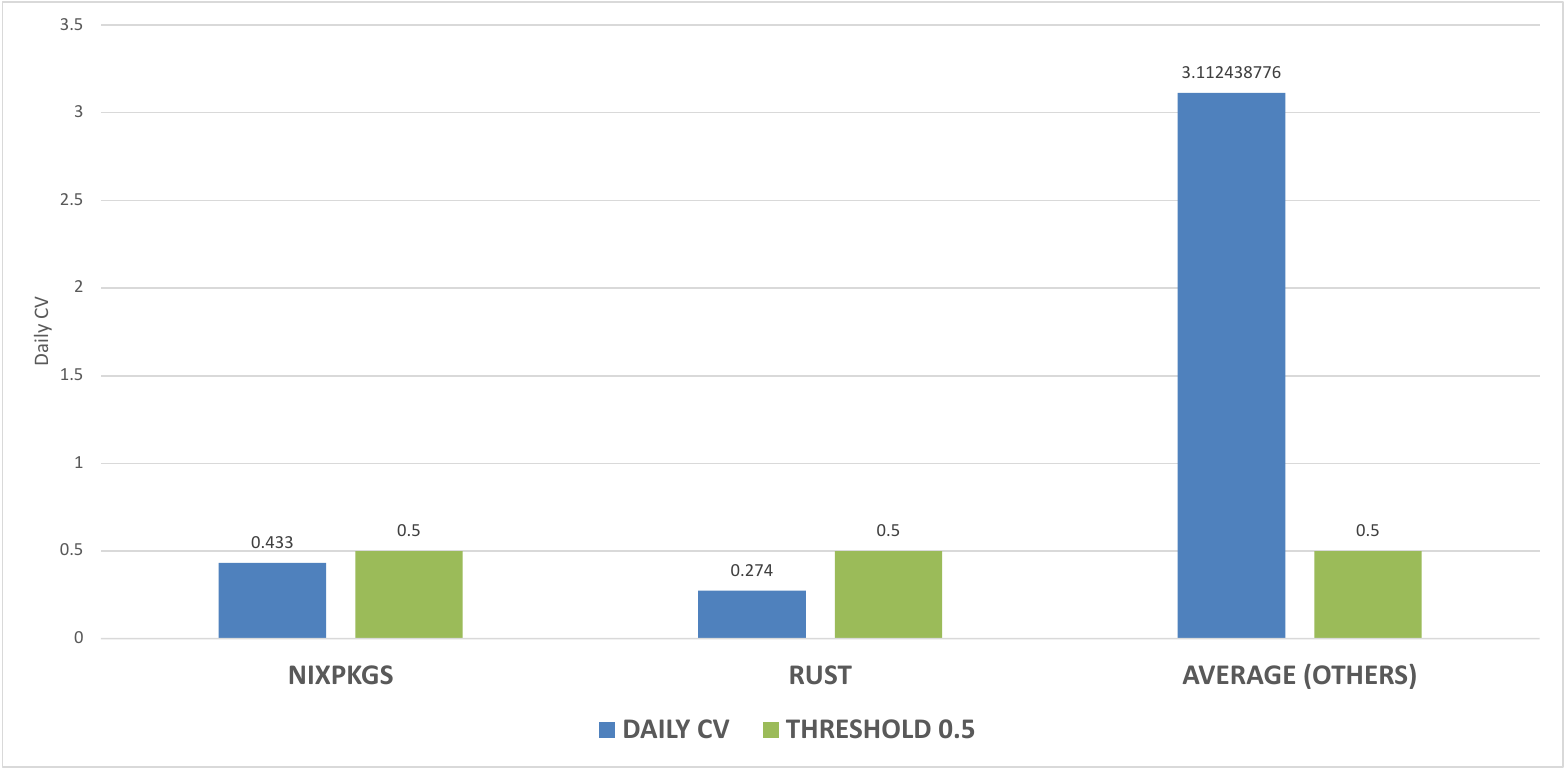}
		\caption{Daily CV: Stable Repos vs Aggregated Others}
        \label{fig:Daily Commit Stability}
\end{figure}

To answer \textbf{RQ$_1$}, we applied the original CSI thresholds to the our dataset of 100 repositories.  
Based on the \textbf{commit frequency} function and daily commit stability threshold defined in equation \eqref{eq:commit_stability} which emphasizes that the daily standard deviation remains less than the mean, only 2\% of our evaluated repositories meet this criteria. The graph \ref{fig:Daily Commit Stability} shows details on the daily commit frequency against the stability threshold.

This finding shows that very few of the projects sustain stable day-to-day commit rhythms without long idle spells or sudden bursts of activity. 
The remaining 98\% exhibit inherently ``bursty'' or ad-hoc behavior. Since only 2 repositories have $\phi_{c} >0$, only these two repositories contribute variance to the CSI calculation. 
This raises concern regarding the suitability and practicality of a daily-level stability criterion for most open-source repositories.

The \textbf{issue resolution} pattern was another important component of the CSI framework. 
Our analysis revealed that 10 repositories from our dataset do not have the issue feature enabled (Django, FFmpeg, Flink, Gitignore, Jenkins, Kafka, Lantern, Laravel, Spark, and WordPress).
We used the remaining 90 repositories for issue resolution analysis.
We found that none of the projects in our dataset achieved the target threshold of $i(t) \geq 0.30$, with the 95th percentile still an order of magnitude lower at 0.018. 
Only three repositories (``laravel/framework'', ``home-assistant/core'', and ``flutter/flutter'') achieved the mean resolution times below the 14-day target. 
However, they still failed to meet the $i(t)$ threshold, demonstrating that quick turnaround alone may be insufficient when the overall closure rate remains low. 

Specifically, because the denominator $N_{i}^{\text{total}}$ in
Eq.~\eqref{eq:issue_resolution_rate} is the cumulative number of issues ever reported, its large magnitude drags the closure ratio toward zero, so most projects do not meet the $i(t)\!\ge\!0.30$ requirement.
This shortcoming stems from \emph{denominator drag}: large, long-lived
repositories accumulate tens of thousands of issues, making the closed issues within the five-year window comparatively negligible.
Furthermore, the median resolution time $\overline{T}_{\text{resolution}}(t)$ is 128 days, but the distribution exhibits a heavy right tail extending beyond four years, inflating the mean to 236 days. Table~\ref{tab:descriptive-stats} shows details of the descriptive statistics across 90 repositories.
Since $\phi_{i} = 0$ for all 90 analyzed repositories, the issue component contributes no variance to the CSI calculation. This suggests that the current thresholds may require adjustment or that alternative normalization approaches should be considered for meaningful statistical inference.

\begin{table}[ht]
\centering
\caption{Descriptive statistics for 90 projects with issues enabled.}
\label{tab:descriptive-stats}
\begin{tabular}{lcc}
\toprule
Statistic & Closure-rate $i(t)$ & Mean resolution age (days) \\
\midrule
Mean & 0.004 & 236 \\
Median & 0.001 & 128 \\
Std. dev. & 0.008 & 238 \\
95th perc. & 0.018 & 670 \\
Min / Max & 0.0 / 0.066 & 3.5 / 1430 \\
\bottomrule
\end{tabular}
\end{table}

Applying the formulation of \textbf{pull request merge rates} in Equation~\eqref{eq:pr_merge_rate} across our 100-repository dataset, we observed that 72\% of the repositories had an average review time of $\overline{T}_{\text{review}}(t) > 5$ days. 
None of the repositories achieved a merge rate $p(t) \geq 0.4$ as defined in Equation~\eqref{eq:pr_stability}. 
This pattern may also be attributed to the denominator drag effect, where long-lived projects accumulate large numbers of pull requests over their lifetime, making the number of merged pull requests within the five-year evaluation window comparatively negligible relative to the historical total.
Since $\phi_{p} = 0$ for all analyzed repositories, the pull request merge rate component contributes no variance to the CSI calculation. 
Table~\ref{tab:descriptive-stats_pr} shows the detailed repository distribution.

\begin{table}[ht]
\centering
\caption{Pull Request Stability across repositories}
\label{tab:descriptive-stats_pr}
\begin{tabular}{lcc}
\toprule
\textbf{Criterion} & \textbf{Pass} & \textbf{Fail} \\
\midrule
Merge-rate $p(t) \geq 0.40$ & 0 & 100 \\
Mean review time $\overline{T}_{\text{review}} \leq 5$ days & 28 & 72 \\
Both criteria simultaneously & 0 & 100 \\
\bottomrule
\end{tabular}
\label{tab:review_threshold}
\end{table}

From the \textbf{activity engagement} stability formulation, the following conditions must be met: 
\begin{equation}
a(t) \geq 0.25 \quad \text{and} \quad \frac{N_{\text{active\_users}}(t)}{N_{\text{total\_users}}(t)} \geq 0.15
\end{equation}
In our analysis, we found that 86\% of our 100-repository dataset met the activity ratio criteria and 95\% met the active user ratio. 
However, only three (3) repositories had $\phi_{a} > 0$.  only these three repositories were within the target ($\mu_{k}$) and tolerance ($\sigma_{k}$) values of 0.35 and 0.10, as shown in Table~\ref{tab:csi-thresholds}. 
These findings suggest that the activity engagement stability index is highly applicable and practicable. 
However, given we only observed variance to the CSI with a few repositories, this suggests that the activity engagement target and tolerance values could benefit from statistical inferences and calibration.

\subsection{Impact of Alternatives on Stability Thresholds (\textbf{RQ$_2$})}

Findings from \textbf{RQ$_1$} reveal the practical application of CSI framework. However, as detailed in Section~\ref{sec:RQ2_methodology}, the CSI framework could benefit from an improved statistical inference applicable to open source software repositories.  
To investigate the impact of variance in \textbf{commit stability} patterns, we utilized a weekly contribution pattern where we defined the threshold $\alpha_{c}$ as:
\begin{equation}
\alpha_{c} = \frac{\sigma_{\text{weekly commits}}}{\mu_{\text{weekly commits}}} \le 0.5.
\label{eq:threhold limit revised}
\end{equation}

\begin{figure}[ht]
		\centering
		\includegraphics[width=0.48\textwidth]{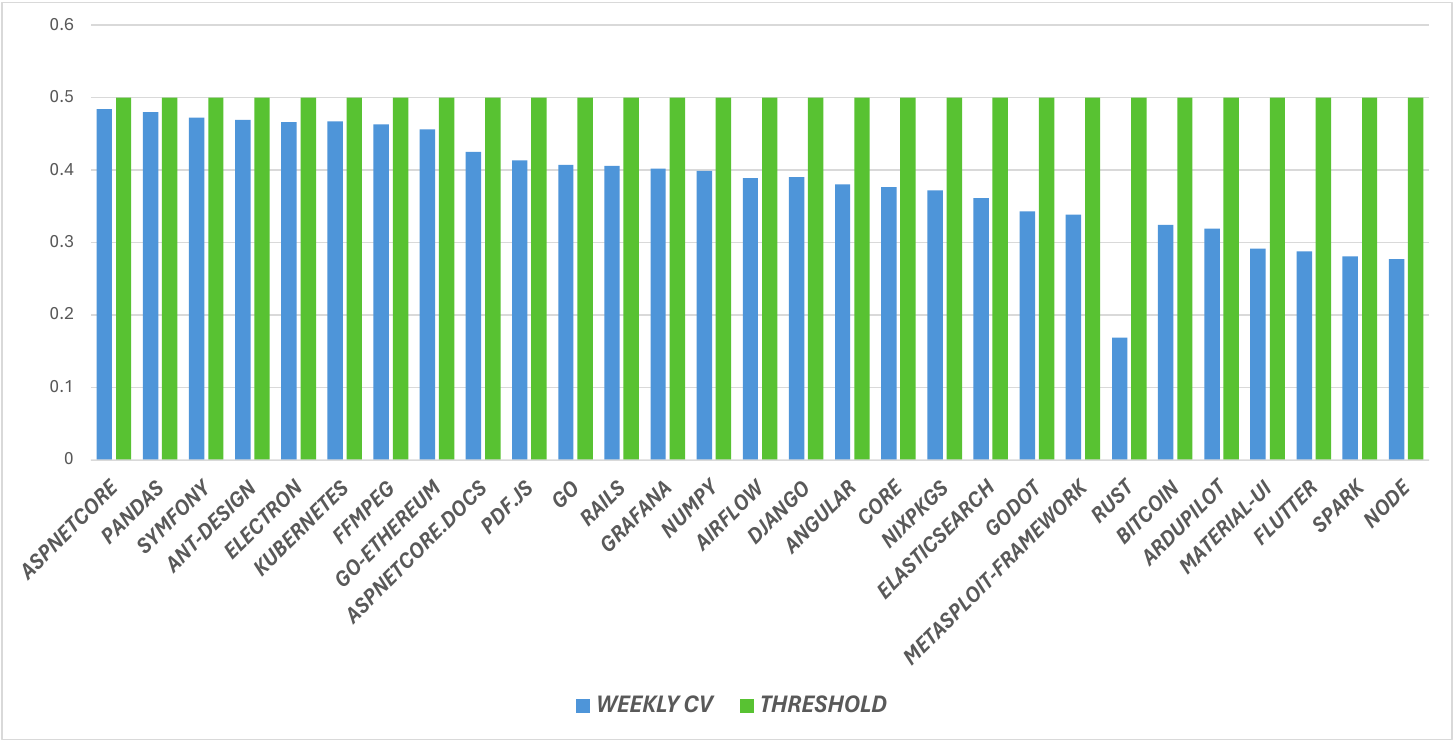}
		\caption{Weekly CV: Stable Repositories}
        \label{fig:Weekly commit_history Stable repositories}
\end{figure}

In our dataset of 100 repositories, we found 29 repositories that had a weekly coefficient of variation (CV) within  the proposed range $\le0.5$. 
This suggests that the weekly aggregation commit frequency pattern may or be more appropriate, as more repositories are able to sustain a genuinely stable week-to-week commit rhythm without long idle weeks or sudden burst of activity in some weeks (over the last 5 years). 
Figure \ref{fig:Weekly commit_history Stable repositories} shows the repositories with stable weekly commit patterns.

From the Table \ref{tab:cv-daily-weekly}, the daily commit counts exhibit more bursts than their
weekly-aggregated counterpart, smoothing to a seven-day window reduces
the median CV by $\approx47\,\%$ and the mean by $\approx52\,\%$
(Table~\ref{tab:cv-daily-weekly}).  
Although the median weekly CV (0.621) still exceeds the CSI stability cut-off of $0.5$, the shift brings a substantial fraction of repositories below the threshold.  

\begin{table}[ht]
\centering
\caption{Central-tendency comparison of commit-frequency
coefficients of variation (CV) at daily versus weekly granularity.}
\label{tab:cv-daily-weekly}
\begin{tabular}{lcc}
\toprule
\textbf{Statistic} & \textbf{Daily CV} & \textbf{Weekly CV} \\ \midrule
Median & 1.182 & 0.621 \\
Mean   & 3.057 & 1.453 \\ \bottomrule
\end{tabular}
\end{table}

Our findings regarding \textbf{issue stability} from \textbf{RQ$_1$} suggests that issue resolution times exhibit significant right-skewed distributions across repositories, with outlier issues requiring substantially longer resolution periods that distort mean-based metrics. 
Furthermore, we observe that the total issue count $N_{i}^{\text{total}}$ significantly influences the $i(t)$ ratio calculation. 
To mitigate the impact of these outliers, we employed the median resolution time $\mathrm{median}\bigl(T_{\text{resolution}}(t)\bigr)$ as a more robust measure. 
This analysis revealed that 64 out of 90 evaluated repositories (excluding 10 repositories with disabled issue tracking) maintain a median issue resolution time of $\le14$ days, indicating relatively efficient issue management practices across the majority of projects. 

To address the dominance of the total issue count denominator, we refined our approach by utilizing ${N_{i}^{\text{total}}\bigl(t,,t+\Delta t\bigr)}$ as the denominator instead. 
This modification constrains the issue closure ratio to the specified time window, yielding a more realistic and temporally-bounded assessment. 
Following these methodological calibrations, we observed that 22 out of 90 repositories were  within the threshold of $i(t) \ge 0.30$. 
However, only 6 repositories had a $\phi_{i}$ score $ > 0$, and only for these repositories did we see contribution to variance in the CSI calculation, as they fell within the target performance ($\mu_{k} = 0.40$) and tolerance bounds ($\sigma_{k} = 0.10$) respectively. Details are shown in Table \ref{tab:descriptive-stats_issue_resolution_refined}

\begin{table}[ht]
\centering
\caption{Descriptive statistics for Issue Resolution across  repositories after calibrations}
\label{tab:descriptive-stats_issue_resolution_refined}
\begin{tabular}{lcc}
\toprule
Statistic & Closure-rate $i(t)$ & Median resolution age (days) \\
\midrule
Mean & 0.233 & 66.800 \\
Median & 0.160 & 5.250 \\
Std. dev. & 0.259 & 199.899 \\
95th perc. & 0.808 & 381.275 \\
Min / Max & 0.001 / 0.967 & 0 / 1331.400 \\
\bottomrule
\end{tabular}
\end{table}

Our findings regarding \textbf{pull request merge rates} from \textbf{RQ$_1$} indicated that the pull request review time exhibit right-skewed distributions across repositories, with some outlier pull request requiring more review times that distort the mean-based metrics. 
We also observed that the total pull request count $N_{p}^{\text{total}}$ influenced the $p(t)$ ratio calculation. 
To mitigate the impact of these outliers we employed the median review time $\mathrm{median}\bigl(T_{\text{review}}(t)\bigr)$ for a more robust measure. 
This analysis revealed that 90\%  of the repositories had the median review time within the 5 days threshold, indicating median review measure is a relatively efficient measure. 
To investigate the dominance of the total pull request count denominator, we refined our approach by utilizing ${N_{p}^{\text{total}}\bigl(t,,t+\Delta t\bigr)}$ as the denominator, i.e total number of pull request within the window being evaluated (5 years). 
Following this modification, we observed that 34\% of the repositories were within the threshold of $p(t) \ge 0.40$. 
However, only 18\% of the repositories had a $\phi_{p}$ score $ > 0$ and fell within the target performance  ($\mu_{k} = 0.50$) and tolerance bounds ($\sigma_{k} = 0.10$).
Therefore, they were the only repositories that contributed variance to the CSI calculation.

\begin{table}[ht]
\centering
\caption{Pull Request Stability across repositories after calibration}
\label{tab:descriptive-stats_pr}
\begin{tabular}{lcc}
\toprule
\textbf{Criterion} & \textbf{Pass} & \textbf{Fail} \\
\midrule
Merge-rate $p(t) \geq 0.40$ & 34 & 66 \\
Median review time $\leq 5$ days & 90 & 10 \\
Both criteria simultaneously & 34 & 66 \\
\bottomrule
\end{tabular}
\label{tab:review_threshold}
\end{table}

As our investigation of \textbf{activity engagement} for \textbf{RQ$_1$} revealed that most repositories met the activity engagement stability threshold hence, no activity engagement stability metric was re-calibrated.

\subsection{Variations in Triangular-Normalizer Parameters (\textbf{RQ$_3$})}

Findings from \textbf{RQ$_2$} provided insights on how statistical inferences can improve the robustness of each component of the CSI. 
This new improvement shifted repositories within the stability thresholds particularly the \textbf{issue resolution}, \textbf{pull request}, and \textbf{activity engagement} measures.  However, for many repositories these still do not contribute to the variance of the $CSI (t)$ because of the thresholds of the proposed variances \eqref{tab:csi-thresholds}. 
Our findings  suggest feasible target and tolerance measures that reward more stable repositories and contribute variance to $CSI(t)$.

For \textbf{issue resolution}, findings from \textbf{RQ$_2$ }, revealed that 22 repositories were within the issue stability threshold. 
However, only 6 repositories had direct impact on the CSI due to the fact that they were not within the target and tolerance measures proposed. 
Building on the empirical evidence from the 22 repositories classified as stable, we determined a more evidence-based target and tolerance $\mu_{k}$ \& $\sigma_{k}$ for issue management $(\phi_{i})$. 
By taking the median and Median Absolute Deviation (MAD) of  $i(t)$ across the stable repositories, we found that target $\mu_{k} =0.620$ and tolerance $\sigma_{k} = 0.221$ are feasible measures rewarding more 10 stable repositories that now contribute variance to the CSI.   
Detailed analysis is shown in Table~\ref{tab:descriptive-stats-Issue_Target&Tolerance}.     

\begin{table}[ht]
\centering
\caption{Descriptive Statistics Issue Resolution Target and Tolerance}
\label{tab:descriptive-stats-Issue_Target&Tolerance}
\begin{tabular}{lr}
\toprule
\textbf{Statistic} & \textbf{Value} \\
\midrule
$\min i(t)$ & 0.3190 \\
$\max i(t)$ & 0.9667 \\
median $i(t)$ & 0.6204 \\
MAD & 0.1489 \\
$1.4826 \times$ MAD & 0.2208 \\
\bottomrule
\end{tabular}
\end{table}

In our analysis of \textbf{pull request merges} for \textbf{RQ$_2$}, we found that 34 repositories were within the pull request merge rate stability threshold. 
However, only 18 of these repositories contributed variance to the $CSI(t)$, again because of the proposed target performance and tolerance bounds of 0.50 and 0.10, respectively. 
Building on the empirical evidence from the 34 stable repositories, we determined a more evidence-based target and tolerance $\mu_{k}$ and $\sigma_{k}$ for pull-request stability $(\phi_{p})$. 
By evaluating the median and Median Absolute Deviation (MAD) of $p(t)$ across the stable repositories, we found target $\mu_{k} = 0.562$ and tolerance $\sigma_{k} = 0.153$ to be feasible measures, resulting in 7 additional repositories being rewarded. These findings match closely with the proposed target and tolerance thresholds for pull-request stability~\ref{tab:csi-thresholds}. 
Details are shown in Table~\ref{tab:descriptive-stats-Pull_Request_Target&Tolerance}.

\begin{table}[ht]
\centering
\caption{Descriptive Statistics Pull Request Merge Rate Target and Tolerance}
\label{tab:descriptive-stats-Pull_Request_Target&Tolerance}
\begin{tabular}{lr}
\toprule
\textbf{Statistic} & \textbf{Value} \\
\midrule
$\min p(t)$ & 0.4167 \\
$\max p(t)$ & 0.9648 \\
median $p(t)$ & 0.5616 \\
MAD & 0.1035 \\
$1.4826 \times$ MAD & 0.1534\\
\bottomrule
\end{tabular}
\end{table}

\textbf{Community Engagement:}
Our findings from \textbf{RQ$_1$} revealed that 86\% of repositories meet the \textbf{community engagement} stability threshold of having a minimum activity ratio threshold of 0.25 and minimum acceptable active user ratio of 0.15. 
However, only three repositories were within the target ($\mu_{k}$) and tolerance ($\sigma_{k}$) values of 0.35 and 0.10, respectively. Building on the empirical evidence from the 86 repositories that met the stability threshold, we evaluated the median and Median Absolute Deviation (MAD) of $a(t)$ across the stable repositories to determine a more evidence-based target and tolerance. 
We found target $\mu_{k} = 3.7056$ and tolerance $\sigma_{k} = 3.2644$ to be feasible measures, which resulted in 64 additional repositories being considered stable contributing variance to $CSI (t) $. 
Details are shown in the Table~\ref{tab:descriptive-stats-Activity_Engagement_Target&Tolerance}.

\begin{table}[ht]
\centering
\caption{Descriptive Statistics Activity Engagement Target and Tolerance}
\label{tab:descriptive-stats-Activity_Engagement_Target&Tolerance}
\begin{tabular}{lr}
\toprule
\textbf{Statistic} & \textbf{Value} \\
\midrule
$\min a(t)$ & 0.3313 \\
$\max a(t)$ & 18.6530 \\
median $a(t)$ & 3.7056 \\
MAD & 2.2018 \\
$1.4826 \times$ MAD & 3.2644\\
\bottomrule
\end{tabular}
\end{table}

\section{Discussion}
\label{sec:Discussion}
Our findings provide novel insights into stability in open source software development, suggesting practical implications for measuring and improving open source project stability.

\subsection{Control Theory \& Stability}

Our empirical analysis across 100 diverse open source repositories provides the first comprehensive validation of the control-theory metaphor for software repositories, examining four major components: commits, issues, pull requests, and community engagement. While previous work has explored the application of control theory to self-adaptive and self-controlling software systems \cite{filieri2015software,kokar1999control}, our findings offer novel practical insights into computing control-theoretic stability within temporal windows in software repositories.

Prior research has extensively studied \textbf{commit frequency} distributions \cite{kolassa2013empirical,alali2008s}, establishing them as a vital component of open source software dynamics. 
However, our findings reveal new insights into the practical derivation of commit frequency stability and demonstrate the applicability of consistent daily and weekly commit rhythms. 
These patterns provide actionable insights for maintainers and administrators of open source projects to better understand repository activity cycles.

Research on issues in open source software has examined various aspects, including user reporting behaviors~\cite{yang2023users} and tools for improved \textbf{issue resolution}~\cite{tao2024magis}. However, to our knowledge, no study has evaluated a practical approach for determining stability in issue resolution within temporal windows. Our findings address this gap and provide valuable insights for maintainers and administrators seeking to optimize issue management processes such as automatically flagging when a project’s closing‐rate drops below a healthy threshold allowing early interventions. Additionally, our analysis reveals a limitation in the overall CSI(t) calculation for repositories with disabled issue features, as these repositories contribute no variance to the overall CSI(t) score.

Previous studies have identified various factors influencing \textbf{pull request acceptance and merging}, including programming language, commit count, and file modifications \cite{soares2015acceptance}. Research has also shown that contributor awareness of integration times motivates continued project participation \cite{gousios2016work}, making pull request timing prediction an area of significant interest \cite{de2021predicting}. Our findings contribute practical insights into stable pull request merge patterns across repositories, with implications for both maintainers seeking to optimize workflows and contributors evaluating potential project engagement based on historical merge patterns.

\textbf{Community engagement} represents a critical factor in open source software success. A comprehensive literature review~\cite{kaur2022understanding} revealed that most studies employ surveys and questionnaires as primary research methodologies, identifying key engagement factors including joining processes, contribution barriers, motivation, retention, and abandonment. Our work provides the first practical, data-driven evaluation of community engagement based on user activity ratios and interaction patterns in issue and pull request comments, offering empirical evidence complementing existing survey-based research approaches.
For example, projects identified via surveys as facing contribution barriers often show sharp drops in \(a(t)\) around the time they introduced new contribution guidelines, providing empirical validation of perceived hurdles. Similarly, repositories with high survey reported retention scores consistently demonstrate stable or increasing \(a(t)\) patterns over time, while projects reporting abandonment issues exhibit declining engagement metrics that precede developer departure by several months. This temporal analysis provide early warning signals for community health issues that traditional survey methods can only detect retrospectively.

\subsection{Supporting Stability Monitoring in Practice}
Existing tools such as CHAOSS \cite{goggins2021open} provide visualization of project health metrics including closed issues, issue age, pull request responsiveness, and new versus repeating contributor ratios, offering valuable insights to maintainers and administrators. Our validated findings on stable commit frequency patterns, issue resolution times, pull request merge rates, and community engagement can be integrated into the CSI(t) metric and incorporated into continuous integration architectures, browser extensions, or project health dashboards. This integration enables teams to access stability metrics that are not only theoretically grounded but also calibrated to real-world repository behavior, which provides evidence that research suggests can enhance developer confidence and trust~\cite{johnson2023make}.
Also, newcomer onboarding can be enhanced by leveraging commit frequency patterns to recommend appropriate tasks. Projects can direct new contributors to actively maintained code areas (indicated by high \(c(t)\) values) where their contributions are more likely to be reviewed and merged successfully.

\section{Threat to Validity}
\label{sec:Threat_to_validity}
We took a number of precautions to reduce the threats to validity of our findings and insights. However, there still remain aspects of our experimental design that could impact the generalizability and validity of our insights. 

\subsection{Repository Size and Selection Constraints}
As defined in Section~\ref{sec:repo_selection}, repositories in our sample dataset were highly ranked and influential. This selection criterion could limit the generalizability of our findings to younger or less popular repositories. However, we deliberately focused on repositories with broader societal impact to ensure our analysis captured stable, well-established projects that represent meaningful patterns in the open source ecosystem.

\subsection{Repository Types}
Prior research \cite{jansen2014measuring} has acknowledged that different repository types (e.g., web applications, libraries, operating systems) may exhibit distinct stability measures due to varying contribution patterns and development practices. We did not categorize repositories by type in this study, which represents a limitation that could be addressed in future work to provide more nuanced stability assessments across different project categories.

\section*{Conclusion}
In this paper, we present findings from an experimental validation of the Composite Stability Index (CSI), a novel control-theoretic framework for measuring software repository stability. Through analysis of 100 highly ranked GitHub projects, we demonstrated that real repositories exhibit equilibrium-seeking behavior across commits, issues, pull requests, and community engagement, mirroring Lyapunov-style stability found in engineered systems. 

Our findings reveal three key insights: First, repositories demonstrate measurable stability patterns that align with control theory principles. Second, weekly commit rhythms provide a more reliable stability measure compared to the daily commit patterns originally proposed in the CSI framework. Third, we derived and recommend a data-driven half-width parameter for the triangular normalizer, which significantly influences CSI variance and improves the framework's practical applicability.

These results not only affirm the practicality of applying a control-theoretic lens to repository stability assessment but also provide the empirical foundation necessary for robust, real-world adoption and improvement of the CSI framework.

\bibliography{references}
\vspace{12pt}
\color{red}

\end{document}